\documentclass[twocolumn,showpacs,preprintnumbers,amsmath,amssymb]{revtex4}
\usepackage{graphicx}% Include figure files
\usepackage{dcolumn}% Align table columns on decimal point
\usepackage{bm}% bold math

\newcommand{\e}{\mathrm{e}}  %e
\newcommand{\msun}{M_\odot}  % Msun
\newcommand{\Mesz}{M\'esz\'aros~}  % \Mesz

\makeatletter
\begin{document}
\preprint{}

\title{High Energy Neutrino Emission from the Earliest Gamma-Ray Bursts}
\author{Shan Gao, Kenji Toma and Peter \Mesz}
\affiliation{Department of Physics, Department of Astronomy and Astrophysics,\\
Center for Particle Astrophysics,\\
The Pennsylvania State University, University Park, 16802, USA}
\date{\today}

\begin{abstract}

We discuss the high energy neutrino emission from gamma-ray bursts resulting
from the earliest generation (`population III') stars forming in the Universe, whose
core collapses into a black hole. These gamma-ray bursts are expected to produce a
highly relativistic, magnetically dominated jet, where protons can be accelerated to
ultra-high energies.  These interact with the photons produced by the jet, leading to
ultra-high energy photo-meson neutrinos as well as secondary leptons and photons. The photon
luminosity and the shock properties, and thus the neutrino spectrum, depend on the mass of the
black holes as well as on the density of the surrounding external gas.  We calculate
the individual source neutrino spectral fluxes and the expected diffuse neutrino
flux for various source parameters and evolution scenarios. Both the individual and
diffuse signals appear detectable in the 1-300 PeV range with current and planned neutrino
detectors such as IceCube and ARIANNA, provided the black hole mass is in excess of 30-100
solar masses. This provides a possible test for the debated mass of the progenitor stellar
objects, as well as a probe for the early cosmological environment and the formation rate
of the earliest structures. 
\end{abstract}

\maketitle

\section{Introduction}
\label{sec:intro}

The first generation of stars in the Universe (known as population III stars)
are the earliest objects to form from the collapse of pristine gas. There is so far no
direct observational evidence for their existence or properties, but simulations 
\cite{abel02,bromm02,yoshida06,Bromm+09} suggest that they are likely to form
with a top-heavy initial mass distribution, with a much heavier average mass than
current stars, in the range $10^2-10^3\msun$, although models with smaller masses are also
possible, e.g. \cite{Norman11}. Such hypermassive stars have very short life times, and except
for a limited intermediate mass range, they undergo a core collapse leading to a
black hole whose mass is some fraction of the initial stellar mass, depending on how much
mass loss occurred during the brief stellar evolution. For sufficiently fast rotating
cores, accretion of remnant gas onto the black hole can lead to the formation of a
powerful jet, resulting in a gamma-ray bursts (GRB), e.g. \cite{Fryer,Heger,Komissarov+09}.

In this paper, we discuss a specific scenario where the GRB jet is initially
dominated by magnetic energy (an MHD jet) \cite{Meszaros+10,Toma+11}. A very large energy
output can be realized in the jet, leading to shocks which can accelerate protons to
highly relativistic energies, leading to significant neutrino emission via the
photomeson process in the presence of the accompanying large photon luminosity. Although difficult
to observe, these neutrinos propagate almost absorption-free across cosmological 
distances, and can provide valuable information about cosmic conditions during
the reionization epoch, when the first structures formed.

In this paper we discuss the very high energy neutrino flux expected from Pop.
III GRBs and the potential for its detectability with current or future large
neutrino detectors such as IceCube \cite{Ic3-rev}, ANITA \cite{ANITA2} and ARIANNA 
\cite{ARIAN-sensit}. In section \ref{sec:model} we discuss the method of
calculation. The main features for our Pop III GRB model and its astrophysical setting are
described in subsection IIa. The photon spectrum serving as the target for the photomeson
process is discussed in subsection IIb. The method used to calculate the photomeson
process and other competing channels for proton and secondary particle energy losses is detailed
in subsection IIc. In section III we discuss the possible Pop III GRB source
rates, and the potential for the detection of the corresponding neutrino fluxes. A summary and
discussion of the implications is given in section IV.

\section{\label{sec:model}Model and Calculation Method}
\subsection{\label{sec:ap}Astrophysical Input}

In the very massive $10^2-10^3\msun$ Population III star scenario, aside from
the approximate $140-260\msun$ range leading to pair instability supernovae which
leave no remnant \cite{Heger}, the core collapse results in a massive black hole (BH) of
mass $M_h$ encompassing a substantial portion of the original mass. A fraction of these
are expected to be fast rotating \cite{Bromm+06,Stacy+11}, one of the requirements for GRB
progenitors. For a star of initial radius $R_\ast$, infall of the remnant stellar gas onto the
black hole leads to an accretion disk of typical outer radius $R_d\sim R_\ast/4$ which for a
typical disk magnetization parameter $\alpha=10^{-1}\alpha_{-1}$ gets swallowed on a
timescale \cite{Komissarov+09}

\begin{equation}
t_{d}\sim\frac{7}{3\alpha}(\frac{R_{d}^{3}}{GM_{h}})^{1/2}
   \sim1.4\times10^{4}\alpha_{-1}^{-1}R_{*,12}^{3/2}M_{h,2.5}^{-1/2}{\rm s},
\end{equation}
where $M_{h,2.5}=10^{2.5}M_{\odot}$ is the largest mass of the central black
hole 
considered here, and henceforth we use the convention $A_{X}\equiv A/10^{X}$.
Extraction of the rotational energy from high mass fast-rotating black holes is
expected predominantly through MHD effects via the Blandford-Znajek mechanism
\cite{bz77}. The corresponding BZ luminosity of the resulting jet is estimated as

\begin{equation}
L_{BZ}\sim\frac{a_{h}^{2}}{128}B_{h}^{2}R_{h}^{2}c .
\label{eq:Lbz}
\end{equation}

Here $a_{h}$ is the dimensionless spin parameter of a Kerr BH , $R_{h}\sim
GM_{h}/c^{2}$ and 
$B_{h}$ is the disk magnetic field strength threading the black hole horizon,
which should 
scale with the disk gas pressure $P$ in the advection-dominated disk (ADAF)
regime 
\cite{Narayan+94}, $B_h^2 = 8\pi P /\beta$, where $\beta = 10\;\beta_1$. Thus,
$B_h \simeq 
(4\sqrt{14}{\dot M}c/3 \alpha \beta R_h^2)^{1/2} \simeq 6.6 \times 10^{13}\;
({\alpha_{-1} 
\beta_1})^{-1} M_{h,2.5}^{-1} M_{d,2.5}^{1/2} t_{d,4}^{-1/2}\;{\rm G}$, where
$M_d$ is the 
accretion disk mass. Assuming a simple scaling $M_d=\delta M_h$, where $\delta
\sim 1$, the 
total energy output of the jet is

\begin{equation}
E_j\simeq L_{BZ} t_d\simeq 2.2\times
10^{55}(a_h^2\delta/\alpha_{-1}\beta_1)M_{h,2.5}~{\rm erg}, 
\label{eq:Ejet}
\end{equation}
and the isotropic equivalent energy is
\begin{equation}
E_{iso} \equiv E_j(1-\cos\theta_j)^{-1}\simeq E_j (2/\theta_j^2)
\label{eq:Eiso}
\end{equation}
where $(1-\cos\theta_{j})\sim \theta_j^2/2$ is the beaming factor for a jet of
collimation half-angle $\theta_j\ll 1$.

The ratio of the photon scattering mean free path and the jet radial dimension
in the comoving frame (the scattering ``optically depth') is very large near the black
hole and drops outward, until it becomes unity at a photosphere located typically beyond
the original stellar radius. As the jet expands beyond this photospheric radius, if the jet
is strongly magnetically dominated, internal shocks are unlikely to occur
\cite{Meszaros+10,Toma+11}, but an an external forward shock (FS) forms as the jet sweeps  up the external
gaseous matter in its surrounding. For an approximately uniform external density of
$n/{\rm cm}^{3}$, at the time $t_d$ when the accretion process stops feeding the jet,
the jet head has reached a distance $r_d$ from the central explosion,

\begin{equation}
r_{d}\sim(\frac{E_{iso} ct_{d}}{4\pi nm_{p}c^{2}})^{1/4}\sim2.8\times
10^{18}E_{57.6}^{1/4}t_{d,4}^{1/4}n_{0}^{-1/4}{\rm cm}
\end{equation}
and the bulk Lorentz factor of the jet head is
\begin{equation}
\Gamma_{d}\sim(\frac{E_{iso}}{4\pi nm_{p}c^{5}t_{d}^{3}})^{1/8}\sim 97
E_{57.6}^{1/8}t_{d,4}^{-3/8}n_{0}^{-1/8}
\end{equation}

e.g. \cite{Meszaros+10}, where from now on we will write $E\equiv E_{iso}$, 
and $m_{p}$ is the proton mass. In the standard GRB  shock description, 
the random magnetic field energy density in the comoving frame of the shocked 
external gas is amplified  to some fraction $\epsilon_B$  of the internal
energy,

\begin{equation}
B\sim(32\pi\epsilon_{B}nm_{p}c^{2})^{1/2}\Gamma_{d}
    \sim3.8 \epsilon_{B,-2}^{1/2}E_{57.6}^{1/8}t_{d,4}^{-3/8}n_{0}^{3/8}{\rm G}.
\end{equation}

In the shock the electrons are Fermi-accelerated into a power law distribution,
resulting in 
synchrotron emission in the above field, and these synchrotron photons are
further subjected 
to scattering by the same electrons leading to a synchrotron-self-Compton (SSC)
radiation 
field \cite{Toma+11}.

If by the time $t_d$ the original jet magnetization parameter has decreased,
which can also be promoted by baryon entrainment from its surroundings, besides the forward shock
(FS) also a reverse shock (RS) will develop, which moves  into the ejecta \cite{zhang+11},
and a hydrodynamical approximation can be used for the description. Let the Lorentz
factor of the unshocked jet in the source frame be $\Gamma_{j}$, the jet head Lorentz factor
be $\Gamma_{d}$ (also measured in the source frame) and the Lorentz factor of the
unshocked ejecta measured in the jet head frame be $\Gamma_{*}$. Since in our case
$\Gamma_{j}$ , 
$\Gamma_{d}\gg1$ , we have

\begin{equation}
\Gamma_{*}\approx\frac{1}{2}(\frac{\Gamma_{d}}{\Gamma_{j}}+\frac{\Gamma_{j}}{\Gamma_{d}}).
\end{equation}
The comoving number density in the unshocked ejecta is

\begin{equation}
n^{\prime}_{j}\approx\frac{E_{iso}}{4\pi m_{p}c^{2}\Gamma_{j}^{2}ct_{d}r_{d}^{2}}.
\end{equation}
The Lorentz factor of the jet as inferred from observed GRB afterglows is 
$10^{2}\sim10^{3}$, and in these calculations we adopt $\Gamma_{j}\approx500$ as a nominal
value. The majority of the reverse shock emission is produced when the reverse shock 
finishes crossing the ejecta and injection of fresh electrons ceases \cite{Murase07}. This 
shock crossing time is essentially $t_d$, at which time the ejecta thickness is approximately 
$ct_{d}$ and the shock radius is approximately $r_{d}$ .

The details of the photon spectrum vary in time and depend on the energy of the GRB jet, the 
shock physics parameters and the details of the environment. The fraction of the shock 
energy that goes into relativistic electrons $\epsilon_e$ (the electron equipartition 
parameter) is assumed to be the same for the forward and reverse shocks;
%so the total electron energy is $2\epsilon_{e}$ ; 
a similar assumption is made for the forward and reverse magnetic field and the accelerated 
proton energy equipartition parameters $\epsilon_{B}$ and $\epsilon_{p}$ (relevant for 
neutrino emission).

The jet total energy $E_j$ is reasonably well defined as a function of the disk mass and the 
BH spin parameter. However, the corresponding isotropic-equivalent energy $E_{iso}$ depends 
on the uncertain jet opening angle. Finally, the radiation produced depends, via the shock 
radius, on the external medium density $n$, which is largely unknown, depending on details 
of the star formation process. For simplicity we treat this density as a constant free 
parameter, our choices being guided by existing numerical simulations of early star 
formation and the results of analyses and modeling of lower redshift GRBs. 
The modeling of observed GRB afterglows suggests that for the $z\lesssim 8$, i.e. later 
generation, so-called Pop. I/II GRBs, the densities of the medium in their environment
typically range over $0.1 < n < 100$ cm$^{-3}$ \cite{Panaitescu+02}. The 
environments of the first stars prior to their collapse has so far only been inferred from 
model numerical simulations, which differ significantly among each other. For example, the 
typical early galactic gas environment could evolve as $n \propto (1+z)^4$ \cite{Ciardi+00}, 
or it might be approximately independent of redshift, $n \sim 0.1$ cm$^{-3}$, as a result of 
stellar radiation feedback \cite{Whalen+04,Alvarez+06}.

The small number of analyses for what are currently the most distant GRBs imply ambient 
densities  for these high redshift GRBs which could be $n \simeq 10^2-10^3 {\rm cm}^{-3}$ 
for GRB 050904 at $z \simeq 6.3$ \cite{Gou+07}, and $n \simeq 1 {\rm cm}^{-3}$ for GRB 
090423 at $z \simeq 8.2$ \cite{Chandra+10}.

As nominal cases, we will discuss mainly the models in Table 1, the most energetic 
examples having $M_h=300\msun$, $(a_h^2 \delta/\alpha_{-1}\beta_1)\sim 1$, $E_j=10^{55.3}$ 
erg, $\epsilon_{B}=0.01$, $\epsilon_{p}=0.1$, jet opening angles $\theta_j=10^{-1},~10^{-2}$ 
and an external density $n=1,~10^2,~10^4$ cm$^{-3}$, the corresponding isotropic equivalent 
energies being given by eq.  (\ref{eq:Eiso}). We also consider smaller black hole masses 
$M_h=100\msun$ and $30\msun$ with correspondingly lower isotropic energies.

\bigskip
\begin{tabular}{|c|c|c|c|c|c|}
\hline 
case & $M_h/\msun$ & $E_{j}/erg$ & $\theta_{-2}$  & $\epsilon_{e,-1}$  & $n/cm^{3}$ \tabularnewline
\hline
\hline 
$A_{300}$& 300 & $10^{55.3}$ & $10$  & $1$  & $1$ \tabularnewline
\hline 
$B_{300}$ & 300 & $10^{55.3}$ & $1$  & $2$  & $10^{2}$ \tabularnewline
\hline 
$B_{100}$ & 100 & $10^{54.8}$ & $1$  & $2$  & $10^{2}$ \tabularnewline
\hline 
$B_{30}$ & 30 & $10^{54.3}$ & $1$  & $2$  & $10^{2}$ \tabularnewline
\hline 
$C_{300}$ & 300 & $10^{55.3}$ & $10$ & $2$ & $10^{4}$ \tabularnewline
\hline 
$D_{300}$ & 300 & $10^{55.3}$ & $1$  & $2$ & $10^{4}$ \tabularnewline
\hline
$D_{100}$ & 100 & $10^{54.8}$ & $1$  & $2$ & $10^{4}$ \tabularnewline
\hline
$D_{30}$ & 30 & $10^{54.3}$ & $1$  & $2$ & $10^{4}$ \tabularnewline
\hline
\end{tabular}

\begin{center} Table 1 \end{center}
\noindent
{\it Table 1 caption:} Pop. III GRB model parameters

\bigskip
\noindent
These parameter values are reasonable, given the various uncertainties, but are by no means 
unique.  Different energy equipartition parameters $\epsilon_{X}$ are possible, and also 
lower Pop. III stellar masses \cite{Norman11}, which would lead  to even lower $M_h$ and 
$E_j$ than those in the table. Such lower values, however, will produce dimmer neutrino 
fluxes (see below) which are not favorable for detection with current or future neutrino 
detectors. Higher $M_h$ (e.g. \cite{Spolyar+09}) and $E_j$ than those  in Table 1 are also 
possible but more speculative.

\subsection{\label{subsec:levelb}Photon spectra}

The GRB photons provide the most abundant targets for the photomeson process (nuclear 
collisions, despite larger cross sections, are much rarer). For times $t\leq t_d$, a common 
feature of all jet models is a photospheric spectral component, arising at a radius $r_{ph}$ 
much smaller than $r_d$, where the photon scattering timescale equals the expansion 
timescale. This component has a quasi-blackbody spectrum\cite{Toma+11}, and its fluence is 
taken to be a fraction $\epsilon_{a}=0.1$ of the jet total energy.

In addition to this quasi-thermal component, the shocks contribute various non-thermal 
photon spectral components, which typically dominate at energies both above and below the 
photospheric component. Electrons are accelerated in the shocks to a power law spectrum 
$dn/d\gamma_{e}\propto\gamma_{e}^{-p}$ for $\gamma_{e}>\gamma_{m}$, where

\begin{equation}
\gamma_{m}\sim\epsilon_{e}\frac{m_{p}}{m_{e}}f(p)(\Gamma_{i}-1)
\end{equation}

is the minimum injected electron Lorentz factor in the shock comoving frame (where $i=*,d$ 
is for reverse and forward shock, respectively), and $f(p)\approx(p-2)/(p-1)$ since 
observations suggest an index $2\lesssim p\lesssim2.5$. Accelerated electrons produce a 
(comoving frame) synchrotron spectrum peaking at

\begin{equation}
E_{m}\approx\frac{3heB}{4\pi m_{e}c}\gamma_{m}^{2},
\end{equation}

and the synchrotron photons are inverse Compton (IC) scattered by the same electrons to 
produce a synchro-self-Compton (SSC) spectrum peaking at

\begin{equation}
E_{m}^{SC}\approx2\gamma_{m}^{2}E_{m}.
\end{equation}

The different parameters of the cases A-D in Table 1  lead to non-thermal photon spectra 
which can be quite different. In case A, the non-thermal photon spectrum consists of the sum 
of the synchrotron emission and the SSC component. In the cases B-D the energy density of 
the photons is significantly higher than in case A. The peak of the original SSC spectrum 
exceeds the threshold energy for two-photon pair formation, $E_{m}^{SC}\gg 
E_{\gamma\gamma}$. Here $E_{\gamma\gamma}$ is the $\gamma\gamma$ self-absorption energy 
above which high energy photons produce cascades of electron-positron pairs. The secondary 
pairs in turn also produce synchrotron emission and IC-scatter photons. Part of the SSC 
spectrum, however, can be suppressed by the Klein-Nishina effect.

In cases B and C, the effects from one generation of cascade pairs are calculated. The 
results suggest that the second generation of pairs affects the spectrum less significantly 
than the first and is neglected for simplicity. In case D the photon energy density is so 
high in the higher mass sub-cases that the copious $\e^\pm$ created by $\gamma\gamma\to 
e^\pm$ can modify the original photon spectrum by multiple Compton scatterings.  The 
compactness parameter (defined as the comoving optical depth to $\gamma\gamma$ effects) can 
be estimated as

\begin{equation}
\ell'=\frac{\sigma_{T}\epsilon_{e}E_{iso}t_{d}^{-1}}{8\pi m_{e}c^{3}\Gamma^{3}r_{d}}
\sim 60 
\end{equation}

for the case $D_{300}$, and about $\ell'\sim 34$ for $D_{100}$, so the spectrum is partly 
thermalized (it would be completely thermalized if $\ell' \gg 10^{2}$, e.g. \cite{Pwx1} ). 
In addition, hadronic cascade photons from neutral pion decay could also affect the spectrum, 
an exact evaluation of such cascades requiring a numerical calculation \cite{Pwx1}. In general, 
however, the effects of hadronic cascades is sub-dominant relative to electromagnetic 
cascades from  $\gamma\gamma$ effects \cite{ZhangMesz01}. To simplify things, in this paper 
we calculate the photon spectrum of case $D$ under two limits: $D(1)$, where the non-thermal 
spectrum is approximated as a summation of the synchrotron and SSC spectrum, and $D(2)$, 
where it is assumed that cascades lead to a completely thermalized (black body) spectrum. 
Due to the uncertainties in case $D$, only the forward shock emission is considered. (As it 
turns out, the neutrino spectra calculated for cases $D(1)$ and $D(2)$ give similar 
predictions for the flux in the ARIANNA energy band of observational interest, since in this 
energy range the pionization efficiency approaches unity; see \S \ref{sec:nuflux}).

The details of these various photon spectra are calculated using the methods described in 
the appendix of \cite{Toma+11}. The resulting photon spectra for all cases in Table 1 were 
calculated. In Figures 1-3 we show only some of the representative cases $A,~B$ and $D$, 
expressed in the observer frame, for the mass sub-case $M_h=300\msun$. The smaller mass 
cases are roughly similar, downscaled versions of these.\\

\subsection{ Proton Acceleration, Cooling \\ and Photomeson Production}

\label{subsec:levelc}

Protons will be Fermi accelerated at the shocks to form a spectrum $N\propto E_{p}^{-2}$, 
e.g. \cite{MeszR}. The acceleration timescale for protons can be estimated as

\begin{equation} 
t_{acc}\sim\eta r_{L}/c=\eta E_{p}/eBc
\end{equation}

where $\eta \sim 1-10$ is a shock-structure dependent coefficient \cite{Waxman01}.

%\begin{equation} \eta\sim\frac{E_{p}}{\sqrt{2}eB\Gamma_{i}c}max[1,(\frac{r_{L}}
%{\sqrt{2}\Gamma_{i}l_{coh}})] end{equation}
%in the relativistic case($i=*,d$ for reverse and forward shock). 
%where $r_{L}$ is the Larmor radius while $l_{coh}$
%is the coherence length of the upstream magnetic field.

The adiabatic cooling time scale is $t_{adb}\sim r_{d}/\Gamma c$ . The accelerated proton is 
also subject to synchrotron cooling with a time scale

\begin{equation}
t_{syn}\sim3m_{p}^{4}c^{3}/4\sigma_{T}m_{e}^{2}E_{p}U_{B}
\end{equation}

where $\sigma_{T}=0.664\mu {\rm barn}$ is the Thomson cross section for electrons and 
$U_{B}=B^{2}/8\pi$ is the magnetic energy density in the comoving frame. The total inverse 
Compton time scale is given by \cite{Jones}.

\begin{equation}
t_{IC}^{-1}=\left(\frac{3 \sigma_T c}{32 \pi \gamma_{p}}\right)
\left(\frac{m_{e}}{m_{p}}\right)^{2} 
m_{p}^{2}c^{4}\int_{0}^{\infty}dEE^{-2}\frac{dn}{dE}
\frac{F(E,\gamma_{p})}{\beta_{p}(\gamma_{p}-1)}
\end{equation}

where $F(E,\gamma)=\gamma[f_{1}(z_{a})-f_{1}(z_{b})]-(E/m_{p}c^{2})[f_{2}(z_{a})-f_{2}(z_{b})]$, 
$z_{a}=(E/m_{p}c^{2}) (\gamma+\sqrt{\gamma^{2}-1})$ , 
$z_{b}=(E/m_{p}c^{2})(\gamma+\sqrt{\gamma^{2}-1})^{-1}$, 
and $f_1(z)$, $f_2(z)$ are functions given in \cite{Jones}.
%$f_{1}(z)=(z+6+3/z)ln(1+2z)-(22z^{3}/3+24z^{2}+18z+4)(1+2z)^{-2}-2+2Li_{2}(-2z)$
%$f_{2}(z)=(z+31/6+5/z+3/2z^{2})ln(1+2z)-(22z^{3}/3+28z^{2}+103z+17+3/z)(1+2z)^{-2}-2+2Li_{2}(-2z)$
%$Li_{2}(z)=\sum_{n=1}^{\infty}z^{n}/n^{2}$ . 
The relativistic protons interact mainly with GRB photons, which are the most abundant 
targets in the acceleration region, including photospheric photons as well as forward and 
reverse shock photons.  When the feeding stops at $t\sim t_d$ the photospheric and reverse 
shock emission decay very rapidly, (unless there is continued central engine outflow),  and 
the forward shock decays as power law afterglow. Thus, the bulk of the photon production and 
proton acceleration peaks around the time $t_d$, and here we concentrate primarily on the 
emission over this timescale.

There are various channels for $p\gamma$ interaction, the two most important ones in our 
case being the photo-meson and the photo-pair channels. In the photo-meson channel, 
neutrinos are mainly produced by the decay of charged pions, Kaons, etc.  For the two 
nominal cases considered in this paper, the ultra-high energy protons have a low 
probability to interact with high energy photons, and Kaon production is not significant 
\footnotemark[0] \footnotetext[0]{(although there may be a small pile up at GZK energies (where charged pions suffer 
significant radiative cooling; while Kaons suffer less), which is beyond the scope of the 
present calculation and would require a Monte Carlo simulation toolkit such as GEANT4 
\cite{Asano+06}).}.

In the photo-meson mechanism approximately equal numbers of charged and neutral 
pions are produced, $p+\gamma\rightarrow p+\pi^{0}$ and $p+\gamma\rightarrow n+\pi^{+}$. The 
neutral pions decay into very high energy photons, while the charged pions lead to neutrinos 
$\pi^{+}\rightarrow\mu^{+}+\nu_{\mu}$ with $\mu^{+}\rightarrow 
e^{+}+\nu_{e}+\bar{\nu}_{\mu}$. In this paper, we use a two-step-function approximation for 
the photo-pion production cross section and for the averaged inelasticity to approximate 
single-pion production and multi-pion production. In this approximation the cross section is

\begin{equation}
\sigma_{\phi\pi}(\epsilon_{r})=\begin{cases}
340\mu {\rm b} & {\rm for}~~\epsilon_{thr}=390<\epsilon_{r}<980\\
120\mu {\rm b} & {\rm for}~~\epsilon_{r}>980\end{cases},
\end{equation}
where $\epsilon_{r}$ is the invariant energy, and the inelasticity  is

\begin{equation}
K_{\phi\pi}(\epsilon_{r})=\begin{cases}
0.2 & {\rm for}~~390<\epsilon_{r}<980\\
0.6 & {\rm for} \epsilon_{r}>980\end{cases}
\end{equation}

\includegraphics[scale=0.75]{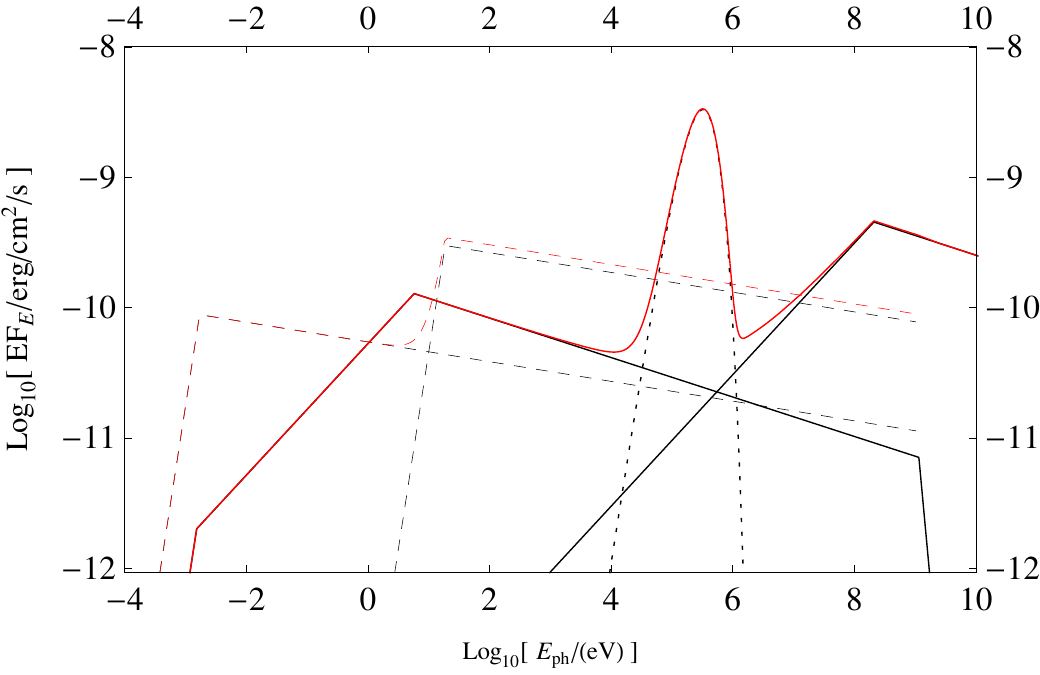}

Figure{[}1{]}: Photon $E^{2}(dN/dE)$ spectrum in the observer frame
from a Pop. III GRB at $z=20$ , case $A_{300}$ (Figs. 1 through 6 
are for $M_h=300\msun$). Solid lines- left: synchrotron
component, right: SSC component, both from the forward shock. Dotted:
photospheric emission. Solid envelope: total of forward shock
plus photospheric emission. Dashed lines- left: synchrotron
component, right: SSC component, both from the reverse shock. Dashed
envelope: total of the reverse shock emission.

\includegraphics[scale=0.85]{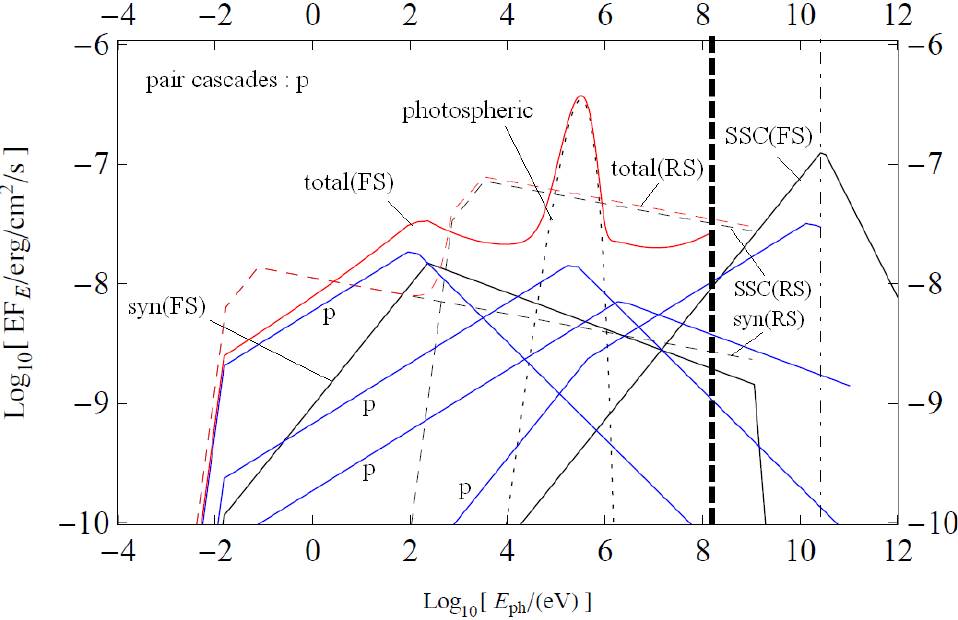}

Figure{[}2{]}: Photon $E^{2}(dN/dE)$ spectrum in the observer frame
from a Pop. III GRB at $z=20$ , case $B_{300}$. Solid lines- left, syn(FS):
synchrotron component, right, SSC(FS): SSC component, both from forward
shock. Dotted: photospheric emission. Other solid lines labeled ``p":
emission from electron-positron pair cascades. Overall solid envelope:
total forward shock emission, included photospheric emission and cascade.
Dashed lines- left: synchrotron component, right: SSC component, both
from reverse shock. Dashed envelope: total reverse shock emission.
Thick vertical dashed line: photon-photon absorption energy. SSC spectrum
(FS) above $E_{KN}$ (Vertical dot-dashed line) is suppressed by the Klein-Nishina
effect.

\includegraphics[scale=0.75]{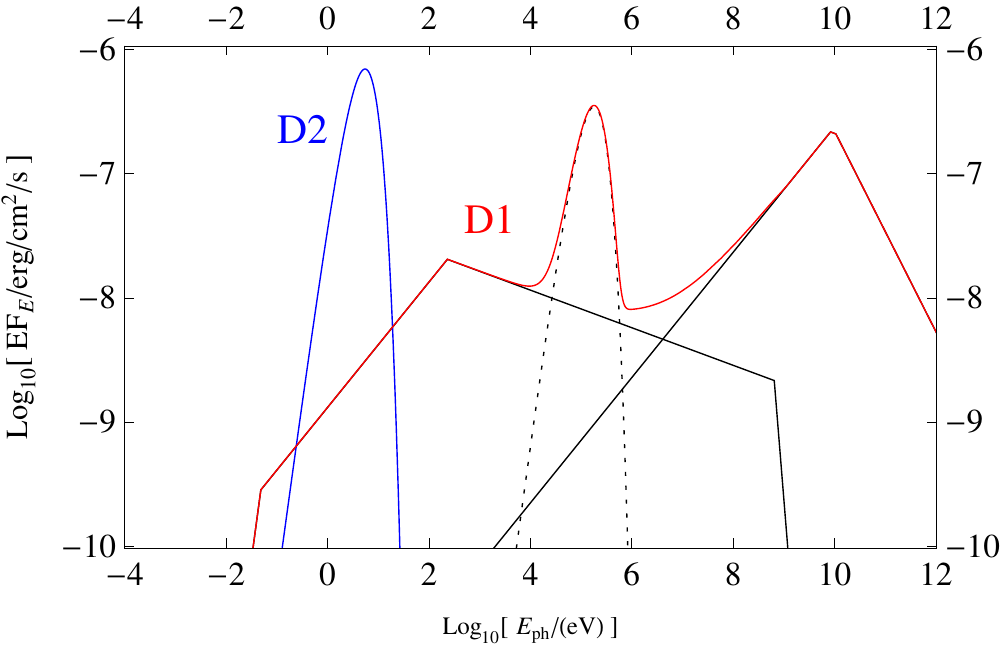}

Figure{[}3{]}: Photon $E^{2}(dN/dE)$ spectrum in the observer frame
of a Pop. III GRB at $z=20$ , case $D_{300}$. Two extreme limits are shown:
Case D(2) assumes complete thermalization of the original spectrum
due to copious pair formation and Compton scatterings; Case
D(1) solid lines show the original synchrotron and SSC from the forward
shock, dashed is the photospheric emission, and the solid envelope
is the sum of these, neglecting spectral changes due to pair formation.

\bigskip
\bigskip
\bigskip

The detailed prescription and the ratio of these products are described in \cite{Dermer+06}. 
The photohadronic cooling time scale is

\begin{equation}
t_{p\gamma}^{-1}=\frac{c}{2\gamma_{p}}\int_{\epsilon_{thr}}^{\infty}d\epsilon\sigma_{\phi\pi}(\epsilon)K_{\phi\pi}(\epsilon)\epsilon\int_{\epsilon/2\gamma_{p}}^{\infty}d\epsilon\epsilon^{-2}dn(\epsilon)/d\epsilon
\end{equation}
where $dn/d\epsilon$ is the photon number density per energy decade in the shock comoving frame.

The other important $p\gamma$ interaction mechanism under the present conditions is photo-pair 
production, $p+\gamma\rightarrow p+e^{+}+e^{-}$. Although in general the energy loss rate 
from this channel is small compared to photo-pion production, due to the fact that the 
average inelasticity per collision is small ( $\sim10^{-2}$ ), in some cases and at some 
energies multiple {interactions} can take place and the energy loss rate can become 
appreciable compared to the photo-pion rate. Here we treat this effect using an approximate 
analytical method described in \cite{Dermer+09}, taking the photo-pair cooling time scale to 
be

\begin{equation}
t_{\phi e}^{-1}(\gamma_{p})\approx \frac{A}{\gamma_{p}^{2}}
\int_{\gamma_{p}^{-1}}^{\infty}d\epsilon\frac{n_{ph}(\epsilon)}{\epsilon^{2}}
\{(2\gamma_{p}\epsilon)^{3/2}
[ln(\frac{2\gamma_{p}\epsilon}{k_{\phi e}})-\frac{2}{3}]+\frac{2}{3}k_{\phi e}^{3/2}\}
\end{equation}

where$A=({7m_{e}\alpha_{f}c\sigma_{T}}/{9\sqrt{2}\pi m_{p}})$ is a constant, 
$\alpha_{f}\approx1/137$ is the fine structure constant and $k_{\phi e}=1$ is a numerical 
constant selected to approximate the relevant cross-section. Charged pions and muons are 
also subject to energy loss due to radiative cooling. The dominant channel is the 
{synchrotron cooling}. Due to this effect the logarithmic slope of the neutrino spectrum 
steepens by 2 units \cite{Waxman01,MeszR} above the muon critical energy $E_{\mu b}$ at which
\begin{equation}
t_{sy,\mu}(E_{\mu b})=t_{dec,\mu}(E_{\mu b}),
\end{equation}
where $t_{sy,\mu}$ is the muon synchrotron cooling time scale and $t_{dec,\mu}$ is the muon 
decay time scale in the comoving frame. The neutrino spectrum is further suppressed above 
the pion critical energy $E_{\pi b}$ where $E_{\pi b}$ is given by $t_{sy,\pi}(E_{\pi 
b})=t_{decay,\pi}(E_{\pi b})$. Relativistic neutrons produced through the channel 
$p+\gamma\rightarrow n+\pi^{+}$ have a much longer decay timescale \cite{PDG} in the 
comoving frame and their effect is neglected in this calculation.
\\
\\
The re-acceleration timescale for pions is longer than the pion decay timescale for all 
cases of interest, while for muons $t_{\mu,acc}<t_{\mu,decay}$ when the magnetic field is 
above $\sim5\ G$ . The latter can be realized in the early phase of the afterglow in our B 
and D case. However, for simplicity, we shall assume here that re-acceleration is 
inefficient for all leptons when calculating the photon and neutrino spectrum. A more 
detailed spectrum including the re-acceleration can be evaluated numerically, as in 
\cite{Murase07}.
\\
\\
The total energy loss rate of the protons is given by $t_{p}^{-1}=\sum t_{i}^{-1}(i={\rm 
all~channels})$ and the photo-pion cooling efficiency is defined through 
$f_{\pi}(\gamma_{p})=t_{p\gamma}^{-1}/t_{p}^{-1}$. This gives the average fraction of energy 
lost to pions from the injected protons at energy $\gamma_{p}$. On average each charged pion 
carries a fraction $\sim0.2$ of the energy of its parent proton and each neutrino carries a 
fraction of $\sim0.05$ (either from pion decay or muon decay). Thus we have
\begin{equation}
J_{\nu}(E_{\nu})=(1/4)f_{\pi}(E_{p})f_{\pi b,\mu b}(E_{p})J_{p}(E_{p})
\end{equation}
where $E_{\nu}\sim0.05E_{p}$ and the flux $J_{X}$ is defined by $J_{X}\equiv 
E_{X}^{2}dN(E_{X})/dE_{X}dt$ . The function $f_{\pi b,\mu b}(E_{p})=Min[1,((E_{\pi}/E_{\pi 
b})^{-2})]\{[(1/2)Min[1,(E_{\mu}/E_{\mu b})^{-2}]+(1/2)Min[1,((E_{\pi}/E_{\pi b})^{-2})]\}$ 
approximates the effects of pion and muon cooling discussed above. The use of the factor 
$1/2$ assumes that half the neutrinos come from charged pion decay and the other half from 
muon decay. Neutrinos with different flavors approximately contribute equally when they 
oscillate \cite{PDG} over cosmological distances although muons are more cooled than pions 
before they decay and may induce a different flavor ratio \cite{Waxman+05}
\\
\\
\\
In figures 4-6  we show the comoving frame cooling timescales for the various proton 
interaction channels, corresponding to photon spectra of Figs. 1-3, for the forward shock 
regions in the cases $A_{300}$, $B_{300}$ and $D(2)_{300}$.

\newpage

\includegraphics[scale=0.6]{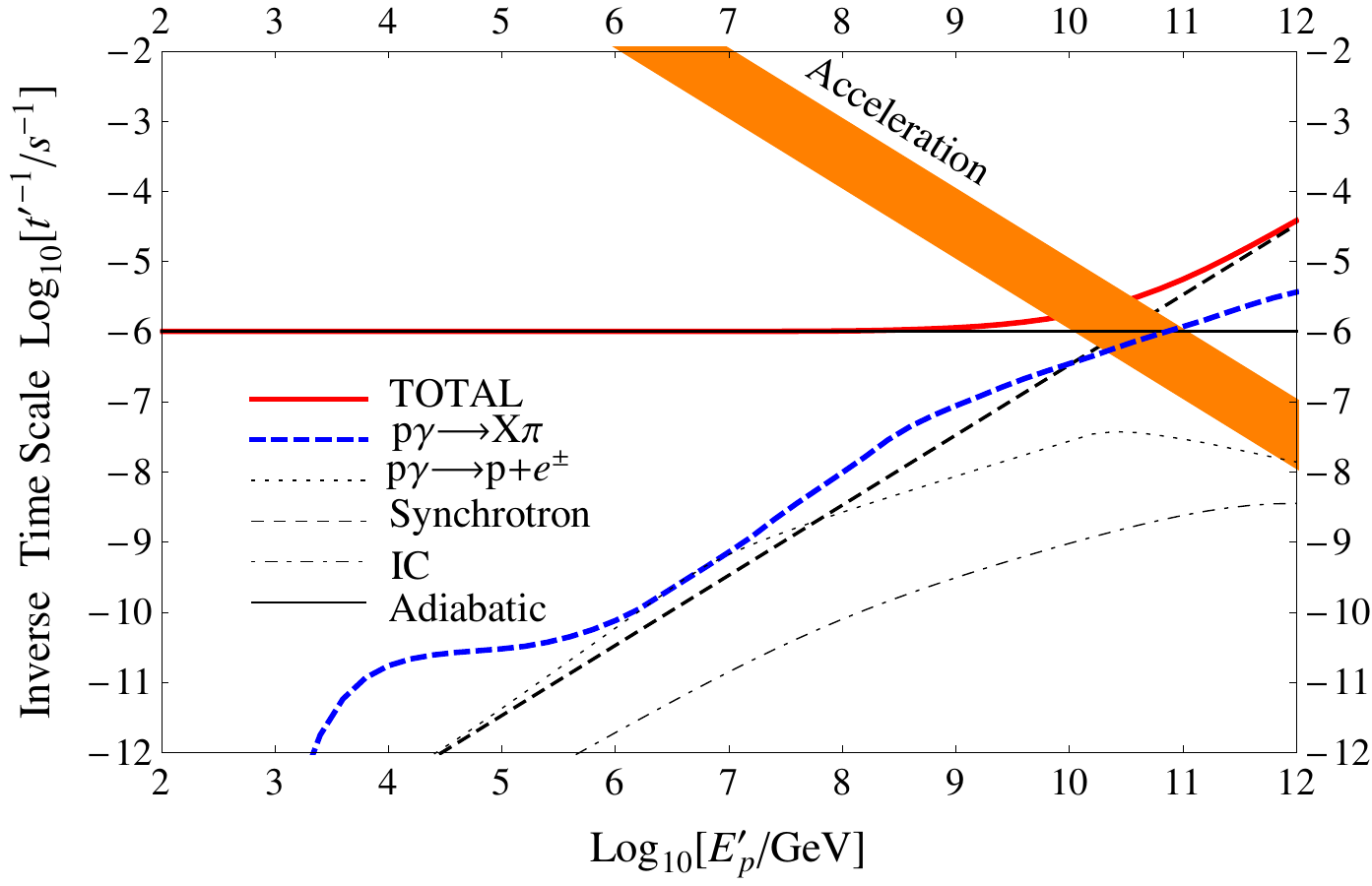}
\\
Figure{[}4{]}: Proton cooling, case $A_{300}$ inverse timescale in the forward shock
region, plotted in the jet comoving frame. The thickness
of the acceleration line shows the uncertainty in acceleration efficiency.

\includegraphics[scale=0.6]{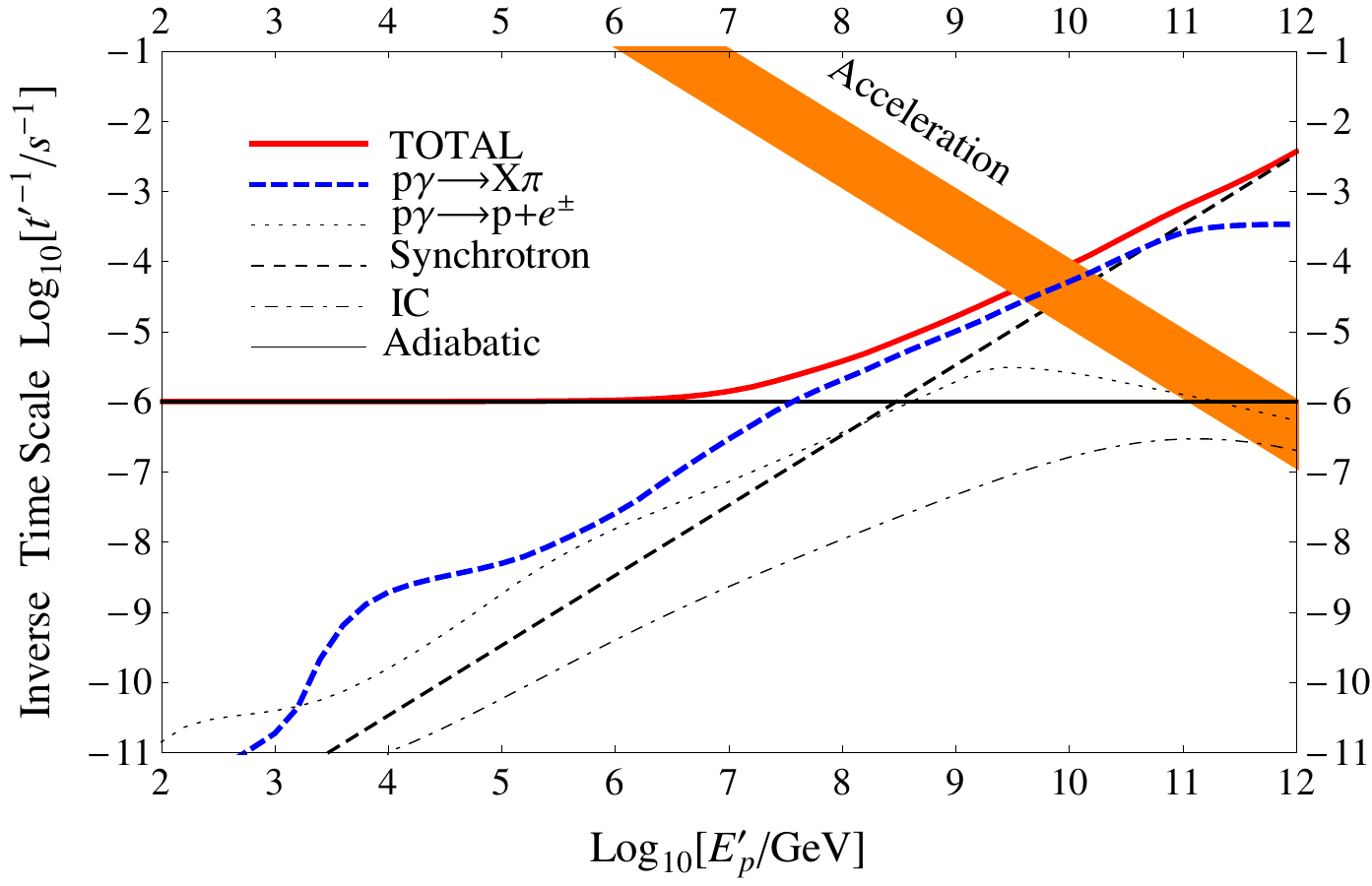}
Figure{[}5{]}: Proton cooling, case $B_{300}$ inverse timescale in the forward shock
region plotted in the jet comoving frame.
\\
\includegraphics[scale=0.6]{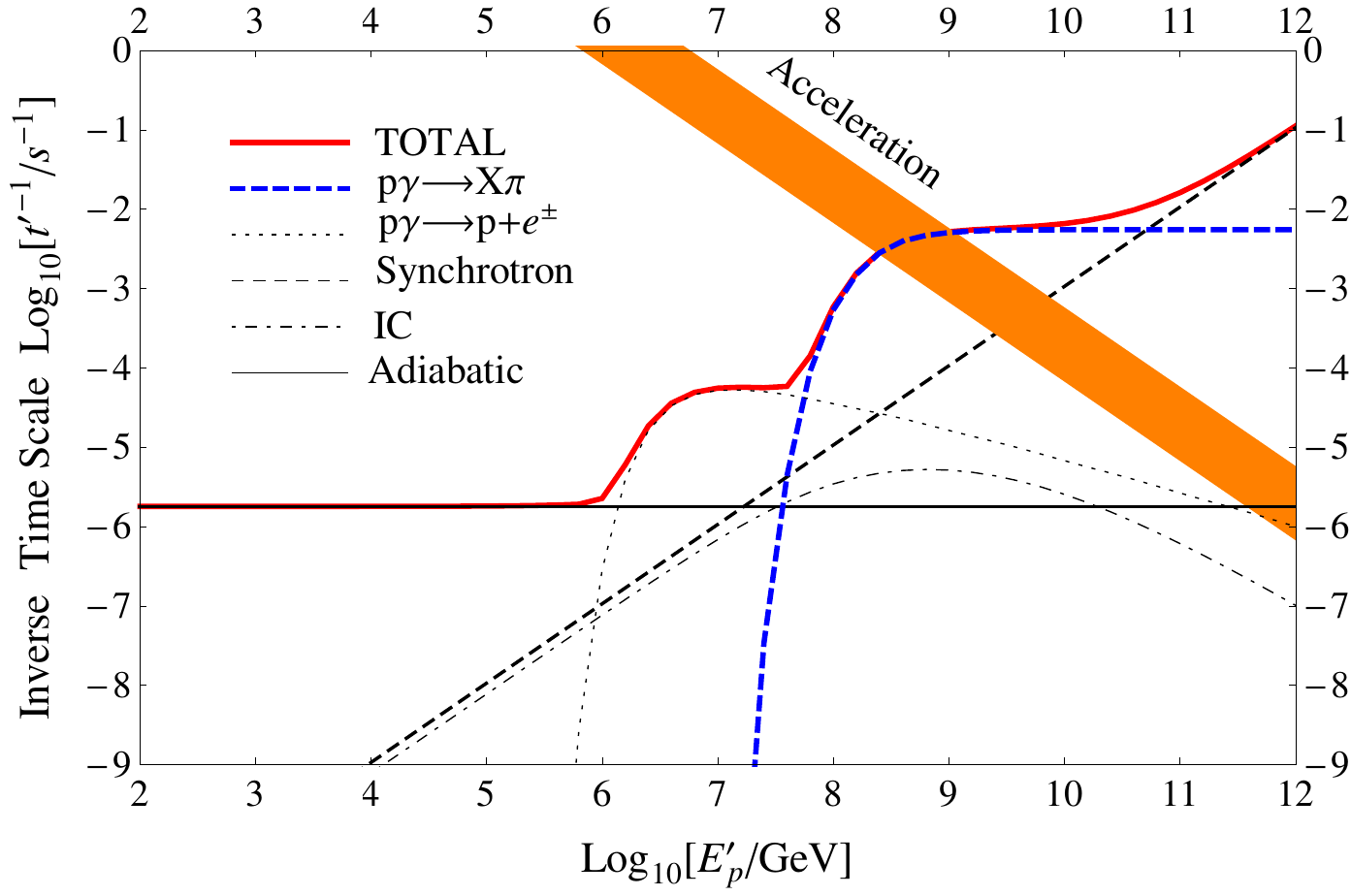}
\\
Figure{[}6{]}: Proton cooling, case $D(2)_{300}$ inverse timescale in the 
forward shock region, in the jet comoving frame. The cutoff in the
dashed line is due to the photomeson threshold and the cutoff in the
dotted line are due to the photo-pair threshold, where the photon energy
in the proton frame are $\sim m_{\pi}\sim140{\rm MeV}$ and $\sim m_{e^{\pm}}\sim1{\rm MeV}$
respectively.

\section{NEUTRINO FLUXES}

\label{sec:nuflux}

\subsection{Individual source neutrino fluence}

\label{subsec:indiv}

Pop. III stars are expected in the re-ionization epoch, at redshifts $z\gtrsim 7$ 
\cite{Trenti10,Loeb+01}, and the majority could arise at redshifts $z\sim 20$, with the very 
first objects at redshifts possibly as high as $z\lesssim 70$ \cite{Naoz}. Here we calculate 
the neutrino spectrum for an individual Pop. III GRB using the model and 
approximations described in sections I and II, for the various cases listed in Table 1 of 
section II. The total neutrino fluence spectra (the time integrated energy spectral flux) 
are evaluated at the deceleration time $t_d$, when the GRB emissivity is the largest. As an 
example, the source fluences for the highest mass cases $M_h=300\msun$ are plotted in Figure 
7 in the observer frame, for the four nominal cases $A_{300}$ through $D(1,2)_{300}$, 
assuming a redshift $z=20$. At low energies the spectrum is dominated by the 
$\Delta$-resonance where the proton energy reaches the threshold energy. At higher energies 
multi-pion processes become significant. The total emission is dominated by the interaction 
with the external shock photons, especially synchrotron photons (due to the original 
electrons and, depending on the case, secondary pairs), for which the photo-pion efficiency 
approaches order of unity $\sim o(1)$ . The IC and SSC photons contribute less significantly 
due to their low number densities. This is the reason why the shape of the total neutrino 
spectrum is similar between the cases $A-C$. However in case $D(2)$, where photons are 
assumed to be complete thermalized, there is a dearth of high energy photons above the Wien 
tail so only very high energy protons meet the photo-pion threshold, and the neutrino 
spectrum is different from those in cases $A-C$.

As seen in Figure 7, the neutrino emission from such Population III MHD-dominated GRB models 
is characterized by a very high peak flux energy, as well as a high fluence. The latter is 
due to the high black hole mass, which implies a high intrinsic luminosity as well as a long 
duration of the external shock peak emissivity phase (which is further lengthened in the 
observer frame by the high redshift). The magnetic field strength at the external shock is 
weak compared to that in internal shocks, so the cooling effect for charged pions becomes 
much less significant than when internal shocks may be important, such as in lower redshift 
GRBs. An internal shock component is not included here, since internal shocks are unlikely 
in strongly MHD dominated GRBs, e.g. \cite{Meszaros+10} (for internal shock neutrino 
emission in hydrodynamical models see, e.g.  \cite{Waxman+97,Iocco05,Iocco08,Murase07,Ahlers+11}).

\includegraphics[scale=0.8]{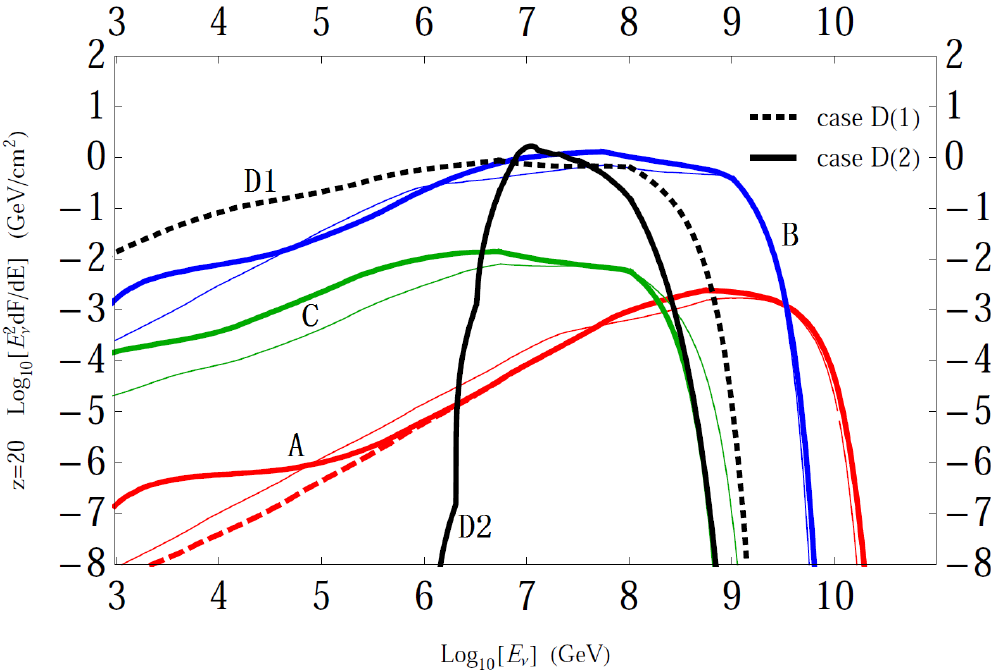}
\\
Figure{[}7{]}: Neutrino fluence (time-integrated energy flux) from one Pop. III 
GRB of $M_h=300\msun$ located at $z=20$ . Model cases $A_{300}$ through $C_{300}$ 
are the labeled solid lines, model $D$ is shown as dashed lines for the
two extreme cases $D(1)$ and $D(2)$ corresponding to the photon spectra
of Fig. 3. The neutrino emission from the forward shocks is shown
by thick lines and that from the reverse shocks as thin lines. The
thick dashed line is the neutrino emission in case A (forward shock)
if photospheric photon emission were absent, to show the effect of
the latter. Cases D(1) and D(2) result in very similar fluence levels
in the multi-PeV range, because the efficiency for photo-meson pion 
production is nearly unity at these energies.
\\

\bigskip
%\begin{center} TABLE 2 goes here \end{center}
\begin{center}
\scalebox{0.8}{%
\begin{tabular}{|c|c|ccc|ccc|ccc|}
\hline
\multicolumn{1}{|c|}{} &  &  & $B$ &  &  & $D(2)$ &  &  & $D(1)$ &
\tabularnewline
\cline{3-11}
$M_{h}$ & $E$ & \multicolumn{1}{c|}{$10$} & \multicolumn{1}{c|}{$20$} & $70$
& \multicolumn{1}{c|}{$10$} & \multicolumn{1}{c|}{$20$} & $70$ &
\multicolumn{1}{c|}{$10$} & \multicolumn{1}{c|}{
$20$} & $70$\tabularnewline
\hline
   & $TeV-PeV$ & $3.0$ & $2.1$ & $1.4$ & $0$ & $0$ & $0$ & $10.3$ & $5.6$ &
$2.1$\tabularnewline
$300$ & $PeV-EeV$ & $6.1$ & $3.2$ & $1.0$ & $2.7$ & $1.5$ & $0.65$ & $3.8$ &
$1.6$ & $0.37$\tabularnewline
   &  &  &  &  &  &  &  &  &  & \tabularnewline
\hline
   & $TeV-PeV$ & $0.71$ & $0.51$ & $0.33$ & $-$ & $-$ & $-$ & $4.5$ & $2.7$ &
$1.1$\tabularnewline
$100$ & $PeV-EeV$ & $1.6$ & $0.83$ & $0.28$ & $0.64$ & $-$ & $-$ & $2.1$ &
$0.90$ & $0.18$\tabularnewline
   &  &  &  &  &  &  &  &  &  & \tabularnewline
\hline
   & $TeV-PeV$ & $0.16$ & $0.11$ & $-$ & $-$ & $-$ & $-$ & $1.8$ & $1.0$ &
$0.40$\tabularnewline
$30$ & $PeV-EeV$ & $0.41$ & $0.22$ & $-$ & $-$ & $-$ & $-$ & $0.72$ & $0.30$
& $0.06$\tabularnewline
   &  &  &  &  &  &  &  &  &  & \tabularnewline
\hline
\end{tabular}}
\end{center}

\noindent
{\it Table 2 Caption}: Number of muon events from an individual burst, in models $B$ and $D$ 
for $M_h=300,~100,~30\msun$ from redshifts $z=10,~20,~70$.
\bigskip

We can obtain an initial quick estimate of the number of muon events expected in a km$^2$
detector from one burst such as, e.g., cases $B$ or $D(1)$, $D(2)$ in Fig. 7. Taking the 
fluence to be $E^2 (dN/dE)\sim 1$ GeV cm$^{-2}$ in all flavors in a band $\Delta E\sim 
10^{6.5}$ GeV around $E\sim 10^{6.5}$ GeV, there are $\Delta N_\nu \sim 3\times 10^{-7}$ 
cm$^{-2}$ neutrinos received over a time $t_{d,obs}\sim 3\times 10^5$ s, or $N_\nu\sim 
3\times 10^3$ km$^{-2}$ per burst. Complete mixing occurs over cosmological distances, so 
1/3 of those are muon neutrinos, and for an approximate conversion probability $P_{\nu\to\mu} 
\sim 1.3\times 10^{-6}(E_\nu /{\rm TeV})$ (valid in the range TeV $\lesssim E_\nu\lesssim$ PeV) 
this translates into $\Delta N_\mu \sim 4$ km$^{-2}$ muon events per burst from a redshift 
$z=20$. A more accurate calculation takes into account that the conversion probability changes 
to  $P_{\nu\to\mu}\sim 10^{-2}(E_\nu/{\rm EeV})^{0.47}$ in the range PeV $\lesssim E_\nu \lesssim$ 
EeV, where EeV$=10^{18}$ eV. Table 2 shows the result of such a calculation performed numerically 
using the detailed spectra of Fig. 7, which gives the number of muon events separately in the 
TeV-PeV and PeV-EeV energy ranges.

The above are only the muon events, in addition to which there would be tau events, 
which can increase the signal (up to at most a factor 2 at the highest energies).
These Pop. III GRB neutrino signals need to be considered against at least three different
sources of background. (a) One is the diffuse atmospheric neutrino background, which has a 
steep spectrum, and at energies $E_\nu\sim 10^{6.5}$ GeV has an upper limit of 
\cite{Ic3-sensit} $E^2 \Phi_E\sim 10^{-9}$ GeV cm$^{-2}$ s$^{-1}$ sr$^{-1}$. Taking an 
angular resolution circle of $\sim 0.7$ degree, this gives over the duration of the burst 
$\Delta N_{\nu,{\rm atm}}\sim 6\times 10^{-5}$ km$^{-2}$ or $\Delta N_{\mu,{\rm atm}}\sim 
2.55\times 10^{-7}$ km$^{-2}$ muons per burst, a negligible background. (b) Another
background is the GZK cosmogenic neutrino background, due to the photo-meson interactions
of the observed ultra-high energy cosmic rays with the cosmic microwave background photons
\cite{Berezinsky+69,Berezinsky-pregal,Engel+01,Yuksel+07,Kotera+10,Waxman11}. 
While the exact value depends on the assumed evolution with redshift of the cosmic ray sources,
it is generally important only at energies higher than those where the signals predicted here 
could be important. (c) A third background is that which may be expected from lower redshift 
(Pop. I/II) GRBs, the typical model for which \cite{Waxman+98,Waxman11} is currently close to being 
constrained by IceCube measurements \cite{Ahlers+11,Abbasi+11}.

From Table 2 one sees that for masses $M_h\sim 300\msun$ at $z=20$ the signal could be a 
doublet or a triplet of events within a $\theta\sim 0.7$ degree error circle within a day,
for models $B_{300}$ and $D_{300}$. In model $D_{300}$ the spectral sub-cases $D(1)$ 
(no pairs) and $D(2)$ (full thermalization due to pairs) bracket the range of possibilities, 
the answer being probably closer to the latter.  For $z\sim 10$ the signals would be larger,
while for $M_h=100\msun$ the signal is a factor $\sim 3-4$ times smaller, and even smaller
for $M_h=30\msun$. For $M_h=30\msun$, a model $B_{30}$ would be very hard or impossible to 
detect, especially against a background of assumed Pop. I/II GRBs; but even for this mass,
the signal might be a doublet from a redshift $z\sim 10$ for $D(1)_{30}$ (note that 
$D(2)_{30}$ is not appropriate since no pair formation is expected for this mass).

\subsection{Population III GRB rates}
\label{subsec:pop3rate}

For calculating the diffuse flux we need to know the burst rate as a function of redshift, 
which is very uncertain due to the lack of observations of confirmed Pop.~III objects of 
any kind.  Population III GRBs are both much rarer and located at higher redshifts than 
the usually considered GRBs from the second and subsequent generations of stars (i.e. 
Pop. I/II GRBs \cite{Yuksel+07,Kistler+09}). If we assume that the Pop.~III GRB rate 
traces the Pop.~III star formation rate (SFR), the observed all-sky GRB 
rate can be parametrized as $(1+z)d\dot{N}_{GRB}^{obs}/dz \equiv 
\phi_{GRB}^{obs}(z)=\phi_{SFR}^{co}(z) \epsilon_{GRB} P_{ph}(z) dV/dz$ 
\cite{Bromm+06,Toma+11}, where $\phi_{SFR}^{co}(z)$ is the Pop. III SFR per unit comoving 
volume (in units of $M_{\odot}\;{\rm yr}^{-1}\;{\rm Mpc}^{-1}$), $\epsilon_{GRB}$ is the 
efficiency of the GRB formation (in units of $M_{\odot}^{-1}$), $P_{ph}(z)$ is the detection 
efficiency of photons for a specific instrument such as the Swift BAT, and $dV/dz$ is the 
comoving volume element of the observed area per unit redshift.  Adopting a specific result 
of the extended Press-Schechter simulation for $\phi_{SFR}^{co}(z)$ and assuming that 
$\epsilon_{GRB} \sim 10^{-8}\;M_{\odot}^{-1}$ (similar to that for the ordinary Pop.~I/II 
cases) and $P_{ph}(z) \sim 0.3$ for $z \sim 10-20$ (which corresponds to the case that the 
luminosity function of the Pop.~III GRBs is similar to the ordinary Pop.~I/II GRBs), one 
obtains $\phi_{GRB}^{obs}(z) \sim 0.5\;{\rm yr}^{-1}$ \cite{Bromm+06} (note that Swift BAT 
only covers $\sim 2\pi/3$ of the sky, and then we use $\epsilon_{GRB}$ and 
$\phi_{GRB}^{obs}(z)$ $\sim 6$ times larger than those in \cite{Bromm+06} to obtain the 
all-sky GRB rate).

However, for such bright bursts as we consider, the photon emission can be generally above 
the threshold of Swift BAT, i.e., $P_{ph}(z) \sim 1$ \cite{Meszaros+10,Toma+11}. Furthermore, the 
Pop. III stars can have a higher GRB formation efficiency $\epsilon_{GRB}$. This may be 
written as $\epsilon_{GRB} = \eta_{beam} \epsilon_{BH} \eta_{env}$, where $\eta_{beam} 
\approx\theta_{j}^{2}/2$ is the jet beaming factor. In cases A and C, $\eta_{beam} = 1/200$ 
and in the cases B and D, $\eta_{beam}=1/20000$. The efficiency for the collapse to lead to 
a central black hole leading to a GRB is parametrized as $\epsilon_{BH}$ (in units of 
$M^{-1}_{\odot}$). Several simulations, e.g., \cite{abel02,bromm02,yoshida06} show that 
Pop.~III  stars are likely to have a top-heavy initial mass function (IMF, i.e. mass 
distribution), instead of a traditional negative index power-law (Salpeter) mass function. 
Stars in the mass range $140\sim260M_{\odot}$ undergo a disruptive pair-instability 
supernova explosion and leave no black hole remnant at all. Given the major uncertainties, 
we take here as a simple approximative example a delta function IMF leading to black holes 
of $M_h=300 M_\odot$, $M_h=100 \msun$ or $30\msun$, so that the black hole formation 
efficiency is assumed to be as high as $\epsilon_{BH}\sim o(1)$, $o(3)$ or $o(10)$ per 
$10^3$ solar masses. The remaining parameter $\eta_{env}$ denotes an efficiency factor 
related to the environment under which the GRB jet can be formed. According to 
\cite{Bromm+06}, a requisite is that the envelope of the star be removed by a binary stellar 
companion in order to let the jet break out, which would suppress the GRB efficiency by a 
further order of magnitude. This condition implicitly assumes the jet durations of order 
30-100 s in the GRB frame known from low redshift observations.  In our model, however, 
the jets last $t_d \gtrsim 10^{4}s$, which is enough to break through even rather large 
stars, such as Pop.~III without significant envelope mass loss. As an effective upper limit, 
we can assume $\eta_{env}\sim o(1)$. The above represents an optimal theoretical case, 
which can be used as an upper limit, $\epsilon_{GRB} \sim 5 \times 10^{-6}\;M_{\odot}^{-1}$ 
for cases $A_{300},~C_{300}$, and $\sim 5 \times 10^{-8}\;M_{\odot}^{-1}$ for cases 
$B_{300},~D_{300}$. 
%For the $M_h=100,~30\msun$ sub-cases, the maximum number of BHs produced is 
%3 times or 10 times larger per solar mass, so we can apply the same $\epsilon_{GRB} \sim 
%5 \times 10^{-6}\;M_{\odot}^{-1}$. 
Taking into account the larger $P_{ph}(z)$, we have a 
theoretical (and highly uncertain) upper limit of $\phi_{GRB} \sim 10\;{\rm yr}^{-1}$ 
for cases $B,D$, and in principle two orders of magnitude higher for cases $A,C$.

We consider now some possible indirect observational constraints on Pop.~III GRB rates, 
since direct observational constraints are so far not available. The afterglow of the 
Pop.~III GRB as well as its prompt emission is bright enough to be detected by Swift and 
by ground based 
optical/near-IR telescopes \cite{Meszaros+10,Toma+11}. On the other hand, the current observed 
GRB rate at $z>6$ is $\sim 0.6\;{\rm yr}^{-1}$, i.e., 3 GRBs (GRB 050904,GRB 080913, and GRB 
090423) during the 5-yr operation of Swift. This may be partly because the optical and near-IR 
observations are more difficult at redshifts $z>6$ due to the Ly$\alpha$ absorption in the 
intergalactic medium. According to the current statistical data \cite{Fynbo09}, only a small 
fraction $\sim 25\%$ of GRBs detected by BAT have redshift determinations, because of bad 
conditions for optical and near-IR observations (not only the Ly$\alpha$ drop-off effect but also 
conditions such as weather as well as dust extinction in the host and our galaxy). Although the 
GRB redshift determination rate may be a function of redshift (i.e., higher redshift GRBs suffer 
stronger Ly$\alpha$ drop-off), we may crudely estimate that the intrinsic GRB rate at $z>6$ 
detected by BAT is $\sim 0.6/0.25 = 2.4\;{\rm yr}^{-1}$. Since the BAT covers only $\sim 2\pi/3$ 
of the sky, the estimate of the GRB rate at $z>6$ from the isotropic sky can be $\sim 2.4 \times 
6 \simeq  14\;{\rm yr}^{-1}$.

Another indirect constraint on the rate of GRBs from the Pop.~III very massive stars may be their 
production of very long and bright radio afterglows \cite{Toma+11}. The predicted durations at 1 
GHz can be typically as long as $200\;{\rm yr}^{-1}$ for $z>10$ radio afterglows arising from 
such GRBs. A comparison of the NVSS  catalog (spanning over 1993-1996) and the FIRST catalog 
(spanning over 1994-2001), which effectively cover $\sim 1/17$ of the sky, did not find any radio 
transient sources which could be due to GRB afterglows with significant flux changes over 
timescales of $\sim 5\;$yr \cite{Lev02,GalYam06}. This indicates that the isotropic rate of 
Pop.~III GRBs is constrained to be $< 17/5 = 3.4\;{\rm yr}^{-1}$. From the above two  rough 
estimates, we adopt in this paper a conservative observational constraint on the massive Pop.~III 
GRB rate of $n_b \lesssim 3 \;{\rm yr}^{-1}$.

\subsection{Diffuse neutrino flux and detectability}

\label{subsec:diffnu}

With the above estimates of the rate of Pop. III GRBs we can now evaluate the diffuse neutrino 
flux expected over a given integration time, for comparison with the capabilities of some current 
or planned neutrino telescopes such as IceCube \cite{Ic3-sensit}, ANITA \cite{ANITA} and ARIANNA 
\cite{ARIAN-sensit}.

In Figure 8 we plot the diffuse neutrino flux from $M_h=30\msun$ Pop. III GRBs, the lowest mass 
case considered, averaged over a period of a year of observation, in units of ${\rm GeV}~ {\rm 
cm}^{-2}~{\rm s}^{-1}~{\rm sr}^{-1}$. This figure assumes, for illustrative purposes, a 
conservative GRB rate of $n_b=1$ observed event per year (taking into account beaming effects), 
and assumes that the bursts occur predominantly at a given redshift (or narrow range of 
redshifts) indicated in the figure, up to an upper limit of $z=70$ for Pop. III formation 
\cite{Naoz}.  The nominal case of one burst per year implies of course an anisotropic flux, but 
for a multi-year integration time, an averaged diffuse flux would be approximated. Also, if the 
rate were larger (e.g. in accord with the above observational indirect limit, say $n_b\sim 3 {\rm 
yr}^{-1}$), the diffuse fluxes would be higher by a factor $n_b$.

The neutrino spectra shown in Fig. 8 are for the cases $B_{30}$ and $D(1)_{30}$ discussed above, 
since these are the preferred candidates for detection. However, for this $M_h=30\msun$ sub-case, 
even from $z=10$ the predicted fluxes just approach the IceCube 5 year sensitivity \cite{Ic3-sensit}
or the ARIANNA 5 year sensitivity \cite{ARIAN-sensit} (if in the latter we extrapolate the
6-month values to 5 years using a $t^{1/2}$ scaling, which could be too conservative if the 
sensitivity is signal limited). For this $M_h=30\msun$ mass we have used the 
spectral case $D(1)$ instead of $D(2)$ because the lower luminosity leads to a $\gamma\gamma$ 
compactness too low to lead to significant thermalization.  The diffuse fluxes for the cases A 
and C are not shown, since they are too low to be of observational interest. The relatively 
larger chance of detection in cases $B$ and $D$ is attributable to the smaller jet opening angle 
and/or a higher external medium density.

In Figure 9 we plot the diffuse neutrino flux from $M_h=100\msun$ Pop. III GRBs, averaged over a 
period of a year of observation, assuming again a GRB rate of $n_b=1$ event per year (including 
beaming effects) and assuming they arise from various redshifts. For this and the higher mass case 
we show the $B$ and $D(2)$ spectral models, the latter assuming that $\gamma\gamma$ effects have 
thermalized the photon spectrum. In this mass case, 5 years of observations would appear to make 
detection feasible if they arise predominantly from redshifts $z\sim 10$, or from $z\sim 20$ for a 
larger $n_b\sim 3$ yr$^{-1}$.

Figure 10 shows the diffuse neutrino fluxes predicted for the highest mass case $M_h=300\msun$, 
averaged over a year of observation for a GRB rate of $n_b=1$ events per year (beaming effects 
included), from various redshifts. The spectral cases shown are again the $B$ and $D(2)$ models, 
which give the highest fluxes (cases A and C giving observationally negligible fluxes, even at this 
high mass \footnotemark[0] \footnotetext[0]{In order 
to reach the minimum detectability with IceCube or ARIANNA, even for 
$M_h=300\msun$ one would need for cases A and C a rate of at least $100~{\rm yr}^{-1}$ if the 
redshift is as low as $z=7$, and an even higher rate if $z$ is higher.}
). For these high mass models, 5 years of observations would make detection feasible even if they arise predominantly 
from redshifts $z\sim 20$ and the rate is as low as $n_b=1$ event per year.

These Pop. III GRB diffuse neutrino fluxes of Figs. 8, 9 and 10, for all three mass values 
and for $n_b=1$ yr$^{-1}$, do not exceed the Waxman-Bahcall cosmic ray limit (including evolution
in redshift \cite{Bahcall+01}, line marked ``WB bound"; see also \cite{Mannheim+01}) or the 
nominal GZK cosmogenic neutrino flux \cite{Berezinsky+69,Engel+01,Waxman11} (line labeled ``GZK"), 
and they are compatible with the observational constraints set by the ANITA-2 mission 
\cite{ANITA2,Vieregg+11}.
The atmospheric neutrino background decreases extremely steep and is only relevant 
here at energies $E_\nu\lesssim 10^{5.5}$ GeV, where its value is $\sim 10^{-9}-10^{-8}$ 
GeV cm$^{-2}$ s$^{-1}$ sr$^{-1}$, e.g. \cite{Ic3-sensit}.

\includegraphics[scale=0.5]{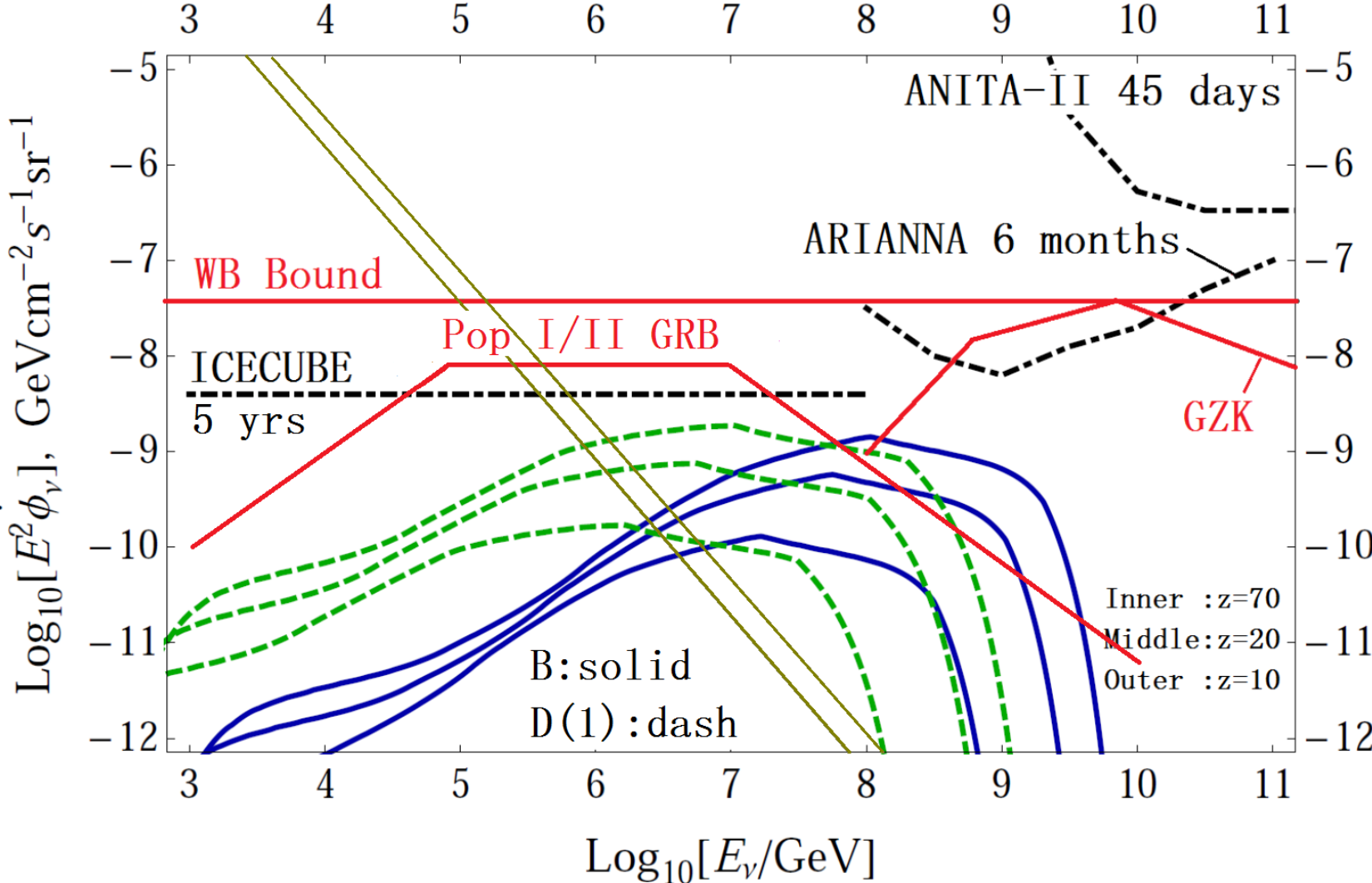}
\\
\\
\\
Figure{[}8{]}: Diffuse neutrino flux in units of GeV cm$^{-2}$ s$^{-1}$sr$^{-1}$
for GRBs of $M_h=30M_{\odot}$, averaged over a year in the observer frame, 
based on the nominal assumption of a rate $n_b=1$ yr$^{-1}$,
for the cases $B_{30}$ and $D(1)_{30}$ (the latter because the compactness parameter is not
large enough as to thermalize the photons to produce a $D(2)_{30}$ case). We have
assumed different typical source redshifts. Also shown are the atmospheric neutrino background,
the IceCube 5 year limits, the ARIANNA 6 month limits and the ANITA II 45 day limits.

\includegraphics[scale=0.5]{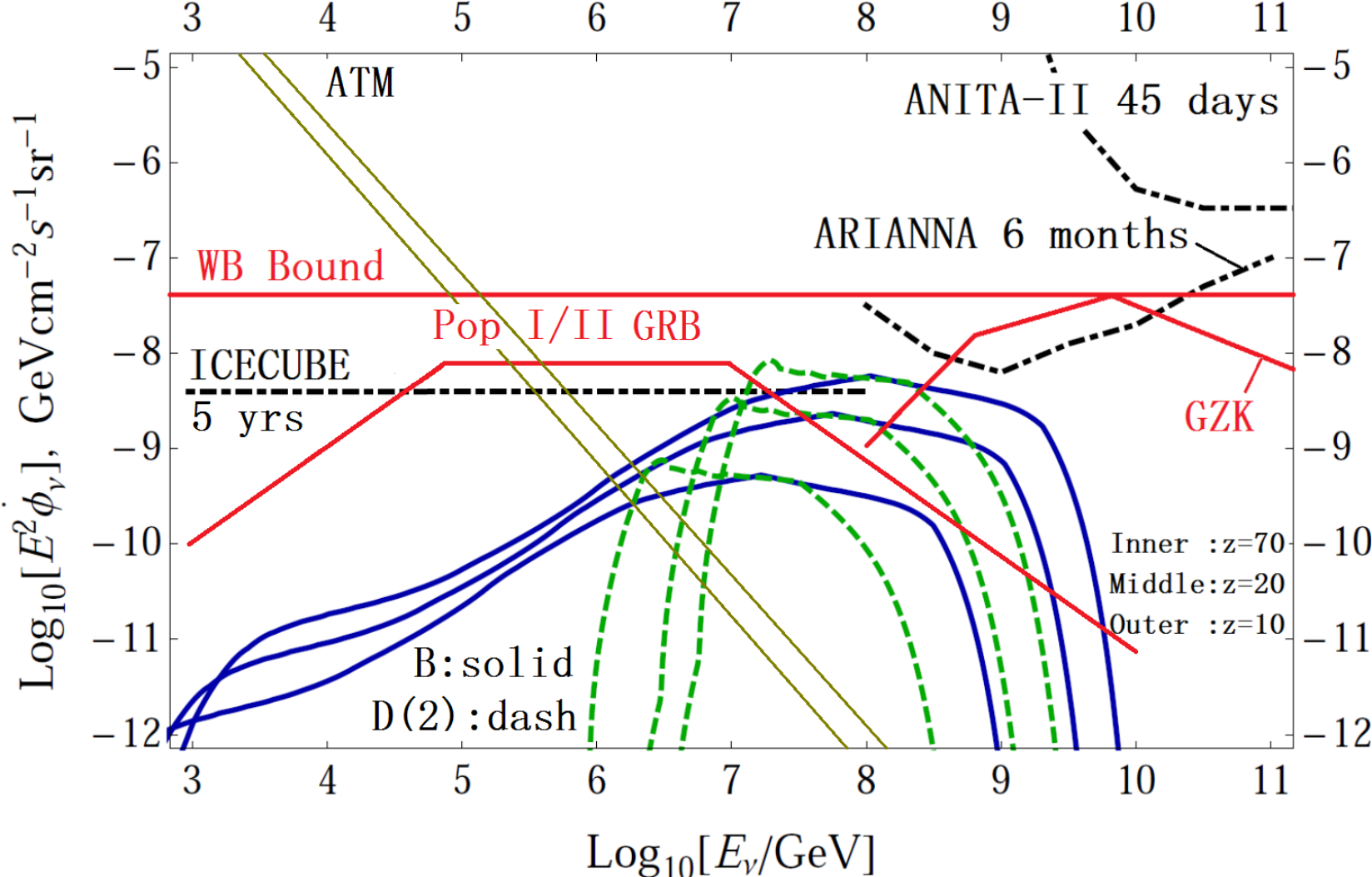}
\\
\\
\\
Figure{[}9{]}: Diffuse neutrino flux in the same units as Fig. 8,
for GRBs of $M_h=100M_{\odot}$, averaged over a year in the observer frame,
based on the nominal assumption of a rate $n_b=1$ yr$^{-1}$,
averaged over a year in the observer frame, for the cases $B_{100}$ and $D(2)_{100}$, 
at different typical source redshifts, with  the atmospheric neutrino background,
the IceCube 5 year limits, the ARIANNA 6 month limits and the ANITA II 45 day limits.
\\
\\
\includegraphics[scale=0.5]{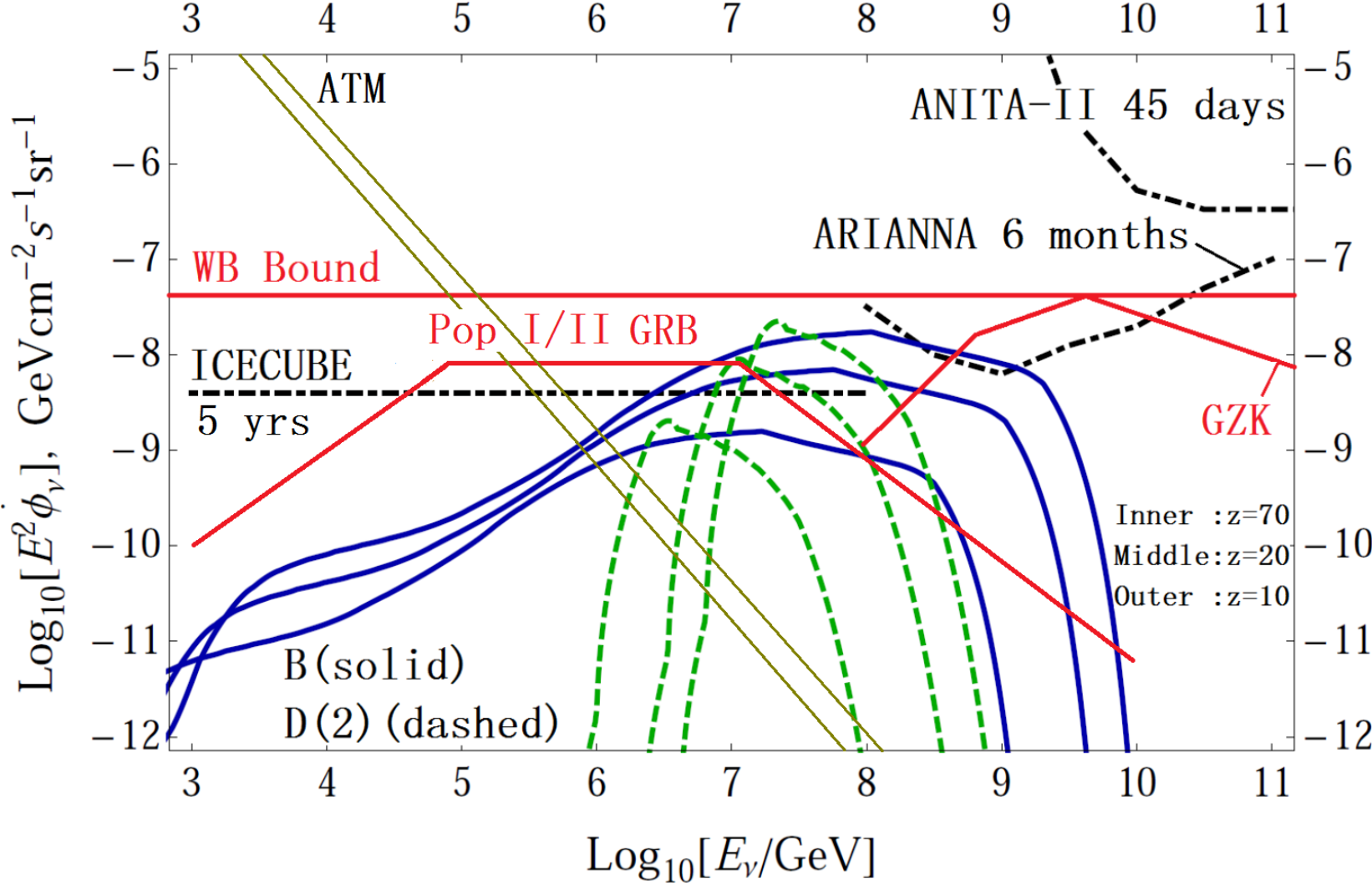}
\\
\\
Figure{[}10{]}: Diffuse neutrino flux in units of GeV cm$^{-2}$ s$^{-1}$sr$^{-1}$,
for GRBs of $M_h=300M_{\odot}$, averaged over a year in the observer frame,
based on the nominal assumption of a rate $n_b=1$ yr$^{-1}$,
for the cases $B_{300}$ and $D(2)_{300}$, at different typical source redshifts, with
the atmospheric neutrino background, the IceCube 5 year limits, the ARIANNA 6 month 
limits and the ANITA II 45 day limits.
\\
\\

\section{Discussion}

\label{sec:disc}

The earliest macroscopic objects to arise out the primeval Universe plasma are the first generation 
Population III stars. These are potentially invaluable probes of the cosmology in the epochs 
$z\sim 10-30$ leading to the currently detected Universe.  The lack of heavy elements at the 
formation of these first objects has led to expectation that they are very massive objects, in 
excess of hundreds of solar masses \cite{abel02,bromm02,yoshida06}, although there is debate about 
the possibility of much lower masses \cite{Norman11}. The almost complete lack of observational data 
about these earliest objects is a major problem. However, objects more massive than about $30\msun$ 
may collapse to become gamma-ray bursts (GRBs) based on what is observationally known at lower 
redshifts, and these could serve as invaluable tracers for this first generation of star formation. 
These GRBs are expected to be bright $\gamma$-ray, X-ray and optical/infrared sources, but 
optical/IR detections providing redshift determinations are largely prevented by the Ly-$\alpha$ 
absorption from the intervening intergalactic gas, while gamma- or X-ray observations would be hard 
to disentangle from those of lower luminosity and lower redshift GRBs. For this reason, predictions 
about the expected high energy neutrino spectrum from Pop. III GRBs, providing a new channel 
completely free of absorption by the intervening medium, could provide an invaluable handle about 
the rate and characteristics of these objects, and through them, about the formation of the earliest 
stars and structures.

One of the largest uncertainties is the initial stellar mass function.  If the Pop. III star masses 
are as large as often assumed \cite{abel02,bromm02,yoshida06,Norman11} the largest black hole masses 
$M_h\sim 10^{2.5}M_\odot$ used here are probably a conservative upper limit to the stellar core 
collapse remnants. The corresponding Pop. III GRB luminosities (eq. (\ref{eq:Lbz})) used here are 
based on the electromagnetic extraction of the rotational energy of a fast rotating black hole, and 
the average beaming angle $\theta_j \sim 10^{-2}$ assumed is comparable to the values  inferred for 
high luminosity Fermi bursts \cite{abdo+09a}. There is no agreement on the value of such parameters; 
for instance, ours differ from those of \cite{suwa-ioka11} who assumed a lower jet luminosity 
(i.e., $a_h^2/\alpha\beta \sim 2 \times 10^{-2}$) and larger jet angle $\theta_j\sim 10^{-1}$ 
calibrated on those of Pop. I/II low redshift GRBs, implying a larger energy input into the 
stellar envelope and less into the emergent jet. On the other hand, the higher jet luminosity 
and narrower jets adopted here lead to a higher ratio of emergent jet to envelope energy.

Other uncertainties concern the rate of occurrence with redshift of Pop. III stars, the resulting 
Pop. III GRB rate, and the external medium density into which the GRB relativistic jet expands.  For 
the formation rates we have based ourselves on recent numerical simulations and theoretical 
estimates (\S \ref{subsec:pop3rate}), moderated by current upper limits on the number of possible 
unidentified high redshift bursts in the currently observed sample, as well as the possible 
contribution of their afterglows to existing transient radio source observations. A conservative 
upper limit that we have considered is one Pop. III burst, on average, per year.  The generally 
considered range of redshifts of occurrence of Pop. III stars is $10\lesssim z \lesssim 30$, with a 
likely typical redshift of $z\sim 20$.  As far as  the external gas densities, we have used $1\leq 
n\leq 10^4{\rm cm}^{-3}$, spanning a range of plausible values. Among the cases considered, those 
with high mass, high luminosity and narrow beam angles lead to high isotropic equivalent energies, 
which together with moderate to high external densities result in relatively high fluxes.

In view of the considerable uncertainty concerning the appropriate parameter values, the predicted fluxes
must be considered as nominal values, to be tested via observations in order to narrow down the range of 
possibilities.

Our calculation of the proton acceleration and the production of high energy neutrinos via 
photo-meson interactions in the GRBs own photon field involve new features not present in previous 
calculations. Typical calculations of the neutrino emission from GRBs, e.g. 
\cite{Waxman+97,Asano+06,Murase07,Becker+10}, considered lower redshift Pop. I/II objects (c.f. 
\cite{Schneider+02}, and generally assumed acceleration in purely hydrodynamic internal shocks (c.f. 
\cite{Dermer+06}). This is also the kind of bursts assumed for the recent IceCube observational 
upper limits \cite{Ic3-40limit}. A basic difference in our case is that Pop. III GRBs are thought to 
involve MHD jets leading to prominent photospheric and external shock emission 
\cite{Komissarov+09,Meszaros+10,Toma+11}, where the target photon field is different. Also, due to 
the higher luminosity the photon density and pair formation effects are more important. This, 
together with a more detailed treatment of the photo-meson cross section and multiplicity results in 
different neutrino spectra.

The prospects for detecting neutrinos from single Pop. III GRBs with IceCube \cite{Ic3-rev} 
(and KM3NeT \cite{KM3NeT} or ARIANNA \cite{ARIANNA}) appear realistic, provided they have
large masses compared to their low redshift counterparts, and provided they are efficient 
proton accelerators. They are also limited the case where the external medium density 
encountered by the jets is high, $n\gtrsim 10^2$ cm$^{-3}$. Such values, although highly 
uncertain, are within the range of what is expected  from numerical simulations. 
As discussed in \S \ref{subsec:indiv}, up to once a year, at $\gtrsim$ 
PeV energies a $300\msun$ GRB at $z\sim 20$ could yield a doublet or a triplet of events 
over a time of a day in IceCube, and in some cases even a $30\msun$ GRB at $z\sim 10$ could 
produce a doublet over a day. In the range 10-300 PeV these signals would be above a diffuse 
background from low-redshift Pop. I/II GRBs, and also of GZK cosmogenic neutrinos of a
different origin (while atmospheric neutrinos are not important at these energies). An accurate 
evaluation of the signal to noise ratio would however depend on model considerations for these 
backgrounds, which is beyond the purpose of this paper. While doublet and triplet searches 
have been done by IceCube over shorter timescales, $\lesssim 100$ s, recent extensions of such
searches to multiplets over $\sim$ day windows are of great interest for the signals discussed 
here.

The detection of single sources would be aided if there were a simultaneous electromagnetic 
detection. The gamma-ray and X-ray flux would in principle be detectable \cite{Meszaros+10,Toma+11}, 
but for the long observer frame durations $t_{d,obs} \sim 10^5$ s $\sim$ 1 day the rise-time is very 
gradual and poses difficulties for normal triggering algorithms \cite{Toma+11}. Optical detections 
are out of the question, due to the blocking by the intergalactic Ly-$\alpha$ absorption
\cite{Lamb+00,Loeb+01}.  However, infrared L-band ($3.4\mu{\rm m}$) or even K-band 
($2.2\mu{\rm m})$ detections may be  possible for some models and redshifts.  For instance, using 
the photon fluxes of Figs. 2 and 3, a rough estimate indicates that for $M_h=300\msun$ at $z=20$, 
neither models $B_{300}$ nor $D(2)_{300}$ are detectable in K but are detectable in L at the level 
of $m_L\sim 3$ and $5$, respectively, which is very bright. Scaling for an $M_h=30\msun$ burst, 
one would expect at $z=20$ neither $B_{30}$ nor $D(1)_{30}$ to be detectable in K, but to be 
detectable in L at $m_L\sim 5$ and $6$ respectively. The L-band at $z=20$ corresponds to 
source-frame UV frequencies, so there could be some intra-source absorption making these dimmer.
Observations of $z \gtrsim 8$ GRBs \cite{Tanvir+09} and galaxy candidates \cite{Bouwen+10} do 
not show evidence for dust, although atomic or molecular resonant scattering could conceivably 
have some effect. Observations in the L-band, while not common, are being done with a few 
telescopes, but these generally have a small field of view, so one-day transients once a year 
such as described above could very easily have gone unnoticed. A detection at these wavebands 
would be dependent on having alerts based on automated triggers from gamma-ray, X-ray or 
neutrino signals.

The prospects for detection of the diffuse neutrino flux are also encouraging, under the
above caveats of large masses, efficient proton acceleration and high external densities.
Using a conservative Pop. III burst rate of $n_b\sim 1$ yr$^{-1}$ and assuming black hole 
mass $M_h\sim 300\msun$ or even $M_h \sim 100\msun$ at $z\lesssim 20$ would lead to a diffuse 
flux, averaged over five years, which at energies $\gtrsim 1$ PeV is within the reach of 
IceCube and ARIANNA, even in the presence of a diffuse neutrino background from lower-redshift 
Pop. I/II GRBs and GZK cosmogenic neutrinos of a different origin. The Pop. III GRB diffuse 
neutrino flux signals have a spectrum which differs significantly from that of the backgrounds 
mentioned above. Thus, if a sufficiently large number of events is accumulated, the spectrum 
should help to distinguish between these signals and the backgrounds. 

We note that the cosmic rays from these Pop. III GRB sources, after $p\gamma$ losses 
within the sources and also in the CMB once outside of them, do not provide a significant 
contribution to the diffuse cosmic ray background. Similarly, the neutrinos they produce
do not contribute significantly to the GZK cosmogenic neutrino fraction.

Thus, in five years or maybe less, IceCube would be able to rule out massive GRBs whose 
formation rate is $n_b\sim 1$ yr$^{-1}$ and $M_h \sim 300\msun$ at redshifts $z\sim 20$, 
or $n_b\sim 3$ yr$^{-1}$ and $M_h\sim 100\msun$ at $z\sim 20$, based on diffuse PeV neutrino 
measurements. The same measurements which are currently setting constraints on the previously
considered Pop. I/II GRB diffuse background \cite{Ahlers+11,Abbasi+11} will also be able to 
set constraints - or perhaps confirm - such models as considered here for the Pop. III GRBs 
and their environment.

Since the neutrino detection of either individual Pop. III GRBs or their diffuse neutrino background 
is only possible for large black hole masses, implying progenitor stars of 
$M_\ast \gtrsim 3 M_h \gtrsim 100\msun$
, large area neutrino experiments at energies $\gtrsim$ TeV would be able to 
address the currently unresolved question of whether Pop. III stars have very large masses or 
perhaps more modest masses approaching solar values. In the latter case, core collapse black holes
and GRBs may not follow from the demise of this first generation of stars; the first black holes 
may arise from subsequent stellar generations, with smaller black hole masses, which would be 
much less luminous than those discussed here. (An ancillary implication would be that a faster 
growth or coalescence rate would be needed to go from such later low mass black holes to the 
supermassive ones inferred in early quasars). In the former case, the presence already in the 
Pop. III era of large mass $\gtrsim 30-100\msun$ black holes resulting in anomalously luminous 
GRB would provide a head-start for the growth into supermassive black holes, as well as provide 
information about the GRB physics in an extreme regime, testing questions of jet physics and 
probing the star forming medium composition and density. Perhaps even more interestingly, 
PeV neutrino measurements could provide  the first positive detections of the earliest  massive 
structures to form in the Universe, at the so far unexplored $10\lesssim z \lesssim 20$ range 
of redshifts.
Such measurements would be invaluable for a better understanding of the 
earliest generation of stars, tracing the cosmic structure formation and the 
environment conditions at the dawn of the Universe.

\bigskip
\begin{acknowledgments}

We are grateful to S. Barwick, D. Cowen, T.DeYoung, D. Fox, D. Grupe, Y. Li, I. Mocioiu
and K. Murase for useful discussions, to the referee for useful comments, and to 
NSF PHY-0757155, NASA NNX 08AL40G for partial support.

\end{acknowledgments}

\appendix

\end{document}